\documentclass[12pt]{article}
\usepackage{graphicx}
\usepackage{marvosym}

\usepackage{mathptmx}       
\usepackage{helvet}         
\usepackage{courier}        
\usepackage{type1cm}        
\usepackage{float}
\usepackage{caption}
\usepackage{amsmath}
\usepackage{longtable}
\usepackage{pbox}

\usepackage{array}
 
\begin{document}

\baselineskip 16pt
\date{}

\title{Qualitative Judgement of Research Impact: 
Domain Taxonomy as a Fundamental Framework for Judgement of the Quality of Research}

\author{Fionn Murtagh (1), Michael Orlov (2), Boris Mirkin (2, 3) \\
(1) School of Computing and Engineering, University of Huddersfield, UK \\
(2) Department of Data Analysis and Machine Intelligence, National \\
Research University Higher School of Economics, Moscow, Russia \\
(3) School of Computer Science and Information Systems, \\
Birkbeck, University of London, UK \\
Email: fmurtagh@acm.org}

\maketitle

\begin{abstract}
The appeal of metric evaluation of research impact has attracted considerable
interest in recent times. Although the public at large and administrative bodies are
much interested in the idea, scientists and other researchers are much more cautious,
insisting that metrics are but an auxiliary instrument to the qualitative peer-based
judgement. The goal of this article is to propose availing of such a well positioned 
construct as domain taxonomy as a tool for directly assessing the scope and quality 
of research.  We first show how taxonomies can be used to analyse the scope and 
perspectives of a set of research projects or papers. Then we proceed to define a 
research team or researcher's rank by those nodes in the hierarchy that have been 
created or significantly transformed by the results of the researcher. An experimental 
test of the approach in the data analysis domain is described.  Although the concept 
of taxonomy seems rather simplistic to describe all the richness of a research domain, 
its changes and use can be made transparent and subject to open discussions.  
\end{abstract}


\section{Introduction: The Problem and Background}

This article constructively supports the view expressed in the Leiden 
Manifesto (Hicks et al., 2015), as well as other recent documents such as 
DORA (Dora, 2013) and 
Metrics Tide Report (Metric Tide, 2016).  All of these advance the principle that  
assessment of research impact should be made primarily according to qualitative 
judgment rather than by using citation and similar metrics. It may be maintained,
due to the lack of comprehensive recording of process, that the 
traditional organisation of qualitative judgment via closed committees is prone to 
bias, mismanagement and corruption. In this work, it is proposed to use 
domain taxonomies for 
development of open, transparent and unbiased frameworks for qualitative judgments. 

In this article, the usefulness of this principled approach is illustrated by, 
first, the issue of 
context based mapping and, second, the issue of assessment of quality of research. 
We propose the direct evaluation of the quality of research, and this principled
approach is innovative.  We also 
demonstrate how it can be deployed by using that part of the hierarchy of the popular 
ACM Classification of Computer Subjects (ACM, 2016)
that relates to data analysis, machine 
learning and data mining. We define a researcher's rank by those nodes in the 
hierarchy that have been created or significantly transformed by the results of the 
researcher.  The approach is experimentally tested by using a sample of leading 
scientists in the data analysis domain. The approach is universal and can be applied 
by research communities in other domains. 

In part 1 of this work, starting with section \ref{part1}, there is the engendering
and refining of taxonomy.  We express it thus to indicate the strong contextual 
basis, and how one faces and addresses, policy and related requirements.
In part 2 of this work, staring with section \ref{part2}, ranking is at issue 
that accounts fully for both quantitative and qualitative performance outcomes.

\section{Review of Research Impact Measurement and Critiques}

The issue of measuring research impact is attracting intense attention of scientists 
because metrics of research impact are being widely used by various administrative 
bodies and by public at large as easy-to-get shortcuts for assessment of comparative 
strengths among scientists, research centres, and universities. This is further boosted 
by the wide availability of digitalized data and, as well, by the fact that research 
nowadays becomes a widespread activity.
The number of citations and such derivatives as Hirsch index are produced by a number 
of organizations including the inventors, currently Thomson Reuters (Thomson Reuters, 2016), Scopus 
and Google. There is increasing pressure to use these or similar indexes in evaluation 
and management of research.  There have been a number of proposals to amend the 
indexes, say, by using less extensive characteristics, such as centrality indexes in the 
inter-citation graphs  or by following only citations in the work of ``lead scientists'' 
(Aragn\'on, 2013). Other proposals deny the usefulness of bibliometrics altogether; some 
propose even such alternative measures as the ``careful socialization and selection of 
scholars, supplemented by periodic self-evaluations and awards'' (Osterloh and Frey, 2014), 
that is, a social- and behavioural-based, administrative, exemplary model.  Other, more practical 
systems, such as the UK Research Assessment Exercise (RAE), now the REF, Research 
Excellence Framework), intends to assess most significant contributions only, and in a 
most informal way, which seems a better option. However, there have been criticisms of 
the RAE-like systems as well: first, in the absence of a citation index, the peer reviews 
are not necessarily  consistent in evaluations (Eisen et al., 2013), and, second, in the long run, 
the system itself seems somewhat short-sighted; it has cut off everything which is out of 
the mainstream (Lee et al., 2013). There have been a number of recent initiatives undertaken by 
scientists themselves such as the San-Francisco Declaration DORA (Dora, 2013), Leiden 
Manifesto (Hicks et al., 2015), The Metrics Tide Report (Metric Tide, 2016). DORA, for example, emphasizes 
that research impact should be scored over all scientific production elements including 
data sets, patents, and codes among others (Dora, 2013).  Altogether, these declarations 
and manifestos claim that citation and other metrics should be used as an auxiliary 
instrument only; the assessment of research quality should be based on ``qualitative 
judgement'' of the research portfolio (Hicks et al., 2015).  Yet there is no clarity on the 
practical implementation of these recommendations.

This article is a further step in this direction.  Any unbiased consideration of metrics 
as well as of other systems for assessment of research impact (Eisen et al., 2013; Lee et al., 2013) 
leads to conclusions that ``qualitative judgment''  should be 
a preferred option (Dora, 2013; Hicks et al., 2015; Metric Tide, 2016). 
This article points out to the concept of domain taxonomy which should be used as a 
main tool in actual organization of assessment of research impact in general and 
quality of research, specifically.

The remainder of this article is organized as follows.  We begin by briefly reviewing
direct and straightforward application of domain taxonomy, for supporting 
qualitative judgement.  Relating to the policy-related work of a national research 
funding agency, and to the editorial work of a journal, these preliminary studies 
were pioneering. 

The third section explains how a domain taxonomy can be used for assessing the quality 
of research. The fourth section provides an experiment in testing the approach 
empirically. The fifth section compares the taxonomic ranking  of our sample of 
scientists with rankings over citation and merit. 

\section{Qualitative, Content-Based Mapping, into which  the Quantitative
Indicators are Mapped}
\label{part1}

In this section and in the next section, we develop taxonomies using sets of 
keywords or selected actionable terms.  It is sought to be, potentially, fully
data-driven.  Levels of resolution in our taxonomy can be easily formed through 
term aggregation.  Mapping the taxonomy, as a tree endowed with an ultrametric, 
to a metric space, when using levels of aggregation, provides an approach 
to having focus (in a general sense, orientation and direction) in the analytics.  

Here we give a first example, in which the taxonomies were generated with the goal 
to provide a tool for open and unbiased qualitative judgment in 
such contexts as research publishing and research funding.
Concept hierarchies can be established by domain experts, and deployed in 
such contexts as research publishing and research funding. 

A short review was carried out of thematic evolution of The Computer Journal, relating 
to 377 papers published between January 2000 through to September 2007.  The construction 
of a concept hierarchy, or ontology, was ``bootstrapped'' from the published articles.
The top level terms, child nodes of the concept tree root, were ``Systems -- Physical'', 
``Data and Information'', and ``Systems -- Logical''.   
Noted was that the category of ``bioinformatics'' did not require further concept 
child nodes.  A limited set of sub-categories was used for ``software engineering'', 
these being ``Design'', ``Education'', and ``Programming languages''.  Under the 
top level category of ``Data and information'', one of the eight child nodes was ``Machine
learning'', and one of its child nodes was ``Plagiarism''.  This was justified by the 
appropriateness of the contents of published work relating to plagiarism.
Once the concept hierarchy was set up, the 377 published articles from the seven years 
under investigation were classified, with mostly two of the taxonomy terms being used
for a given article.  There was a maximum of four taxonomy terms, and a minimum of one.
Table \ref{table000} displays the concept hierarchy that was used at that time.  

\begin{table}
\begin{scriptsize} 
\begin{verbatim}
1. Systems -- Physical
   1.1. Architecture, Hardware 
        1.1.1. Networks, Mobile
        1.1.2. Memory
   1.2. Distributed Systems 
        1.2.1. System Modelling
        1.2.2. Networks, Mobile 
        1.2.3. Grid, P2P
        1.2.4. DS Algorithms
        1.2.5. Semantic Web
        1.2.6. Sensor Networks
   1.3. Networks, Mobile
        1.3.1. Mobile Computing
        1.3.2. Networks
        1.3.3. Search, Retrieval
   1.4. Information Delivery
        1.4.1. Energy
               1.4.1.1. Photonics-based
               1.4.1.2. Nano-based
        1.4.2. Displays
        1.4.3. Bio-Engineering Applications
        1.4.4.  Miscellaneous Applications of Materials 
2. Data and Information
   2.1. Storage
        2.1.1. Databases
        2.1.2. Graphics
        2.1.3. Imaging, Video
        2.1.4. Memory Algorithms
        2.1.5. Non-Memory Storage Algorithms
        2.1.6. Network Storage Algorithms
   2.2. Knowledge Engineering
        2.2.1. Data Mining
        2.2.2. Machine Learning
        2.2.3. Search, Retrieval
   2.3. Data Mining 
        2.3.1. Imaging, Video
        2.3.2. Semantic Web
        2.3.3. Complexity 
   2.4. Machine Learning
        2.4.1. Databases
        2.4.2. ML Algorithms
        2.4.3. Reasoning
        2.4.4. Representation
   2.5. Quantum Processing 
   2.6. Algorithms 
        2.6.1. Coding, Compression, Graphs, Strings, Trees
   2.7. Bioinformatics  
   2.8. Computation Modelling
3. Systems -- Logical
   3.1. Information Security
        3.1.1. Networks, Mobile 
   3.2. Software Engineering 
        3.2.1. Design
        3.2.2. Education
        3.2.3. Programming Languages
   3.3. System Modelling
        3.3.1. Software Engineering
        3.3.2. Testing
        3.3.3. Ubiquitous Computing 
        3.3.4. Workflow
        3.3.5. Games
        3.3.6. Human Factors
        3.3.7. Virtual Materials Science
\end{verbatim}
\end{scriptsize} 
\caption{Concept hierarchy, incrementally constructed, representing a view of appropriate 
subject headings for articles published in the Computer Journal,  2000--2007.}
\label{table000}
\end{table}

\begin{figure}[t]
\begin{center}
\includegraphics[width=10cm]{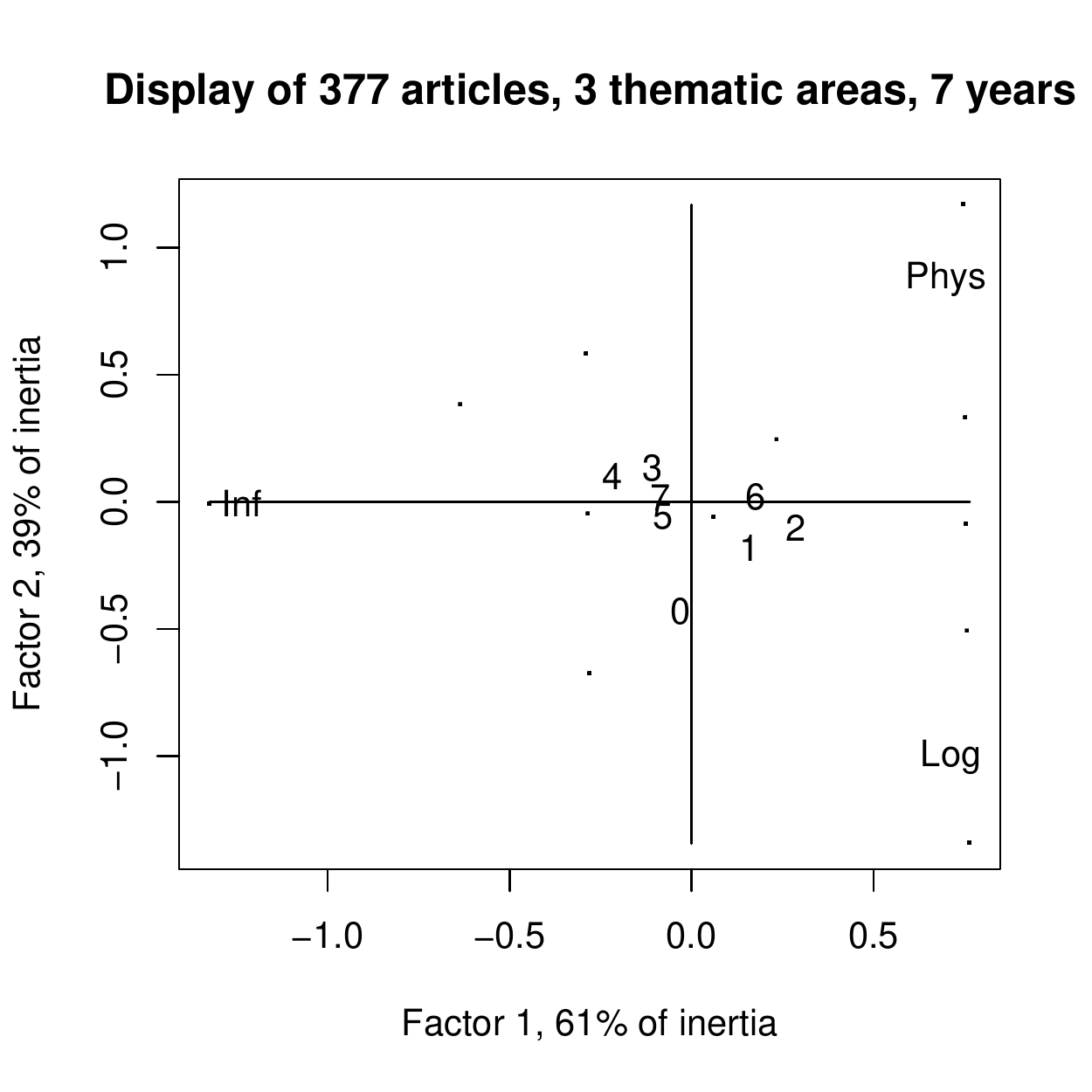}
\end{center}
\caption{Principal factor plane of Correspondence Analysis of 377 published articles 
(positions shown with dots, not all in this central region), crossed by three primary 
thematic areas.  These are: Information and Data (Inf), Systems--Physical (Phys), and 
Systems--Logical (Log).  The years of publication shown (0 = 2000, 1 = 2001, etc.), 
used as supplementary elements in the analysis.} 
\label{fig1}
\end{figure}

A Correspondence Analysis of this data, here with a focus on the top level themes, 
presents an interesting and revealing view.  A triangle pattern is to be seen, 
in Figure \ref{fig1}, where Inf is counterposed on the first factor to the two other,
more traditional Computer Science themes. Factor 2 counterposes the physical and 
the logical
in the set of published research work.  The information displayed in Figure \ref{fig1} 
comprises all information, that is the inertia of the cloud of publications, and of the 
cloud of these top level themes.  The year of publication, as a supplementary 
attribute of 
the publications, is inactive in the factor space definition, and each is projected into 
the factor space.  We see the movement from year to year, in terms of the top level 
themes.
There is further general discussion in Murtagh (2008). 

The perspective described, for archival, scholarly journal publishing, relates to 
the narrative or thematic evolution of research outcomes.  

\section{Application of Narrative Analysis to Science and Engineering Policy}
\label{sect3}

This same perspective as described in the previous section was prototyped for the 
narrative ensuing from national science research funding.  The aim here was thematic 
balance and evolution.  Therefore it was complementary to the operational measures of 
performance -- numbers of publications, patents, PhDs, company start-ups, etc.  In 
Murtagh (2010), the full set of large research centres (8 of these, with up to 
20 million euro funding) and a class of less large research centres (12 of these, 
each with 7.5 million euro funding) were mapped into a Euclidean metric endowed, 
Correspondence Analysis, factor space.  In this space there is displayed the centres, 
their themes, and, as a prototyping study, just one attribute of the research centres, 
their research budget.  The first factor clearly 
counterposed centres for biosciences to centres for telecoms, computing and 
nanotechnology.
The second factor clearly counterposed centres for computing and telecoms to 
nanotechnology. This is further elaborated in section \ref{sect31}.

All in all, there is enormous scope for insight and understanding, that starts from 
subject matter and content.  Quantitative indicators are well accommodated, with their
additional or complementary information.  It may well be hoped that in the future, 
qualitative, content-based analytics, coupled with quantitative indicators, will 
be extended.  For this purpose, it may well be very useful to consider not just 
published research, but all written, and subsequently submitted, research results
and/or plans.  Similarly for research funding, the content-based mapping and assessment 
of rejected work is relevant too, not least in order to contextualize the content of all 
domains and disciplines.  

The role of taxonomy is central to the information focusing that is under 
discussion in this section.  Information focusing is carried out through 
mapping the ontology, or concept hierarchy, as a level of aggregation, corresponding 
therefore to non-terminal, i.e.\ non-singleton, nodes.  
Our interest in this data is to have implications of this for data mining with 
decision policy support in view.  

Consider a fairly typical funded research project, and its phases up to and beyond
the funding decision.  A narrative can always be obtained, in one form or another, and 
is likely to be a requirement. 
All stages of the proposal and successful project life cycle, 
including external evaluation and internal decision making, are highly
document -- and as a consequence narrative -- based.  

As a first step, let us look at the very 
general role of narrative in national research development.   
The following comprise our motivation: Overall view, i.e.\ overall synthesis of 
information; Orientation of strands of development; Their tempo and rhythm.

Through such an analysis of narrative, among the issues to be addressed are the 
following: Strategy and its implementation in terms of themes and subthemes represented;
Thematic focus and coverage; Organisational clustering; 
Evaluation of outputs in a global context; All the above over time.

The aim here is to view the ``big picture''.  It is also to incorporate 
contextual attributes.  These may be the varied performance measures of success 
that are applied, such as publications,
patents, licences, numbers of PhDs completed, company start-ups, and so on.
It is instead to appreciate the broader configuration and 
orientation, and to determine the most salient aspects underlying the 
data.

\subsection{Assessing Coverage and Completeness}
\label{sect31}

SFI Centres for Science, Engineering and Technology (CSETs) are 
campus-industry partnerships typically funded at up to \EUR 20
million over 5 years.  Strategic Research Clusters (SRCs) are also 
research consortia, with industrial partners and over 5 years are
typically funded at up to \EUR 7.5 million.  

We cross-tabulated 8 CSETs and 12 SRCs by a range of 65 terms derived from 
title and summary information; together with budget, numbers of PIs
(Principal Investigators), Co-Is (Co-Investigators), and PhDs.
We can display any or all of this information on a common map, for 
visual convenience a planar display, using Correspondence Analysis.

In mapping SFI CSETs and SRCs, we will now show how 
Correspondence Analysis is based on the upper (near root) 
part of an ontology or concept hierarchy.  This we view as 
{\em information focusing}.
Correspondence Analysis provides simultaneous representation of 
observations and attributes.
Retrospectively, we can project other observations or attributes 
into the factor space: these are supplementary observations or attributes.
A 2-dimensional or planar view is likely to be a gross approximation of 
the full cloud of observations or of attributes.
We may accept such an approximation as rewarding and informative.  
Another way to address this same issue is as follows. 
We define a small number of aggregates of either observations or 
attributes, and carry out the analysis on them.  We then project the 
full set of observations and attributes into the factor space.
For mapping of SFI CSETs and SRCs a simple algebra of themes as set out in 
the next paragraph achieves this goal.  The upshot is that the 2-dimensional
or planar view is a better fit to the full cloud of observations or of
attributes.  

From CSET or SRC characterization as: Physical Systems (Phys), Logical 
Systems (Log), Body/Individual, Health/Collective, and Data \& Information 
(Data), the following thematic areas were defined.

\begin{enumerate}
\item eSciences = Logical Systems, Data \& Information
\item Biosciences = Body/Individual, Health/Collective
\item Medical = Body/Individual, Health/Collective, Physical Systems
\item ICT = Physical Systems, Logical Systems, Data \& Information 
\item eMedical = Body/Individual, Health/Collective, Logical Systems
\item eBiosciences = Body/Individual, Health/Collective, Data \& Information 
\end{enumerate}

This categorization scheme can be viewed as the upper level of a 
concept hierarchy.  It can be contrasted with the somewhat more 
detailed scheme that we used for analysis of articles in the Computer Journal, 
(Murtagh, 2008).  

CSETs labelled in the Figures are: 
APC, Alimentary Pharmabiotic Centre; BDI, Biomedical Diagnostics Institute; 
CRANN, Centre for Research on 
Adaptive Nanostructures and Nanodevices; CTVR, Centre for Telecommunications 
Value-Chain Research; DERI, Digital Enterprise Research Institute; 
LERO, Irish Software Engineering Research Centre; NGL, Centre for 
Next Generation Localization; and REMEDI, Regenerative Medicine Institute.

\begin{figure}
\begin{center}
\includegraphics[width=8cm]{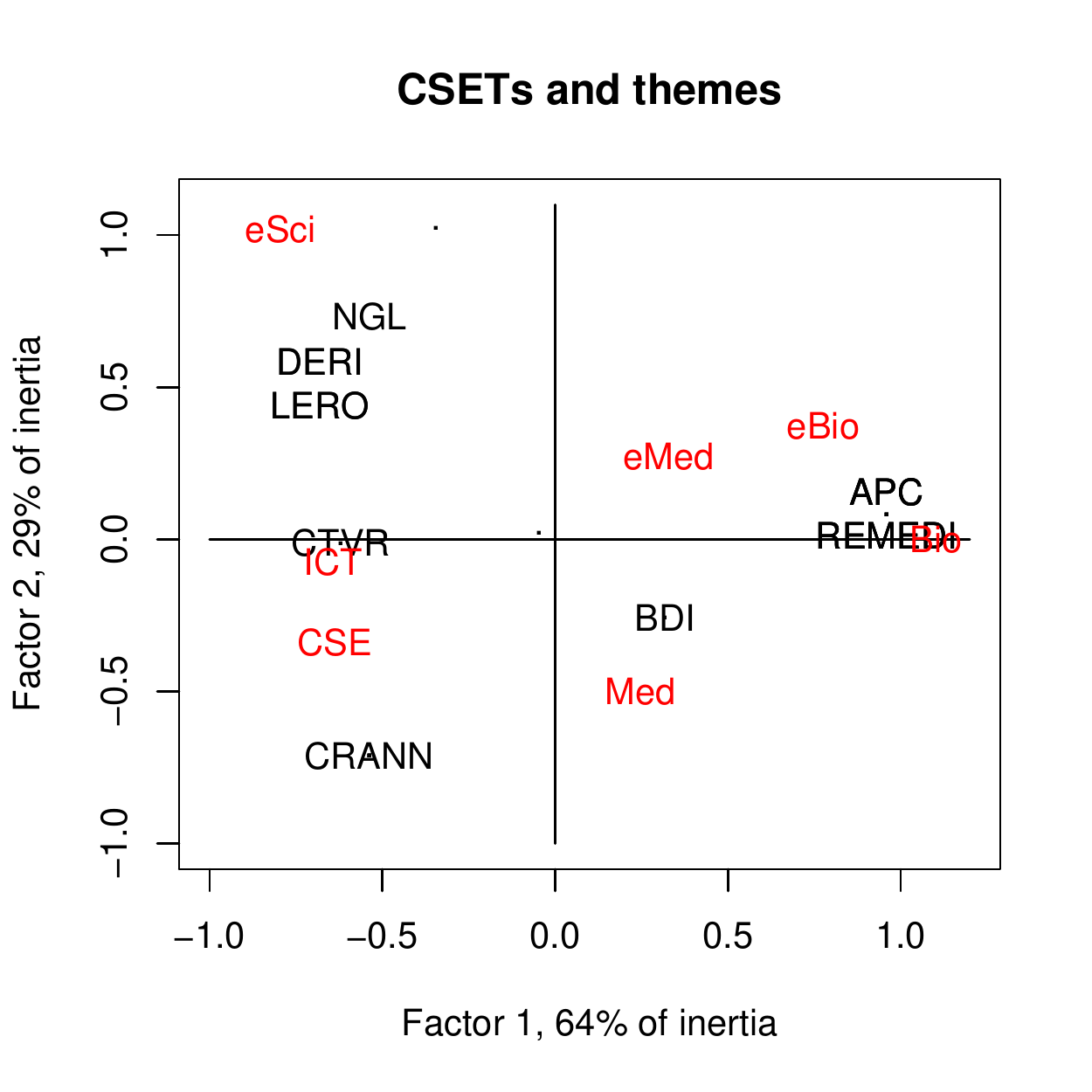}
\end{center}
\caption{CSETs, labelled, with themes located on a planar display, 
which is nearly complete in terms of information content.}
\label{fig9}
\end{figure}

\begin{figure}
\begin{center}
\includegraphics[width=8cm]{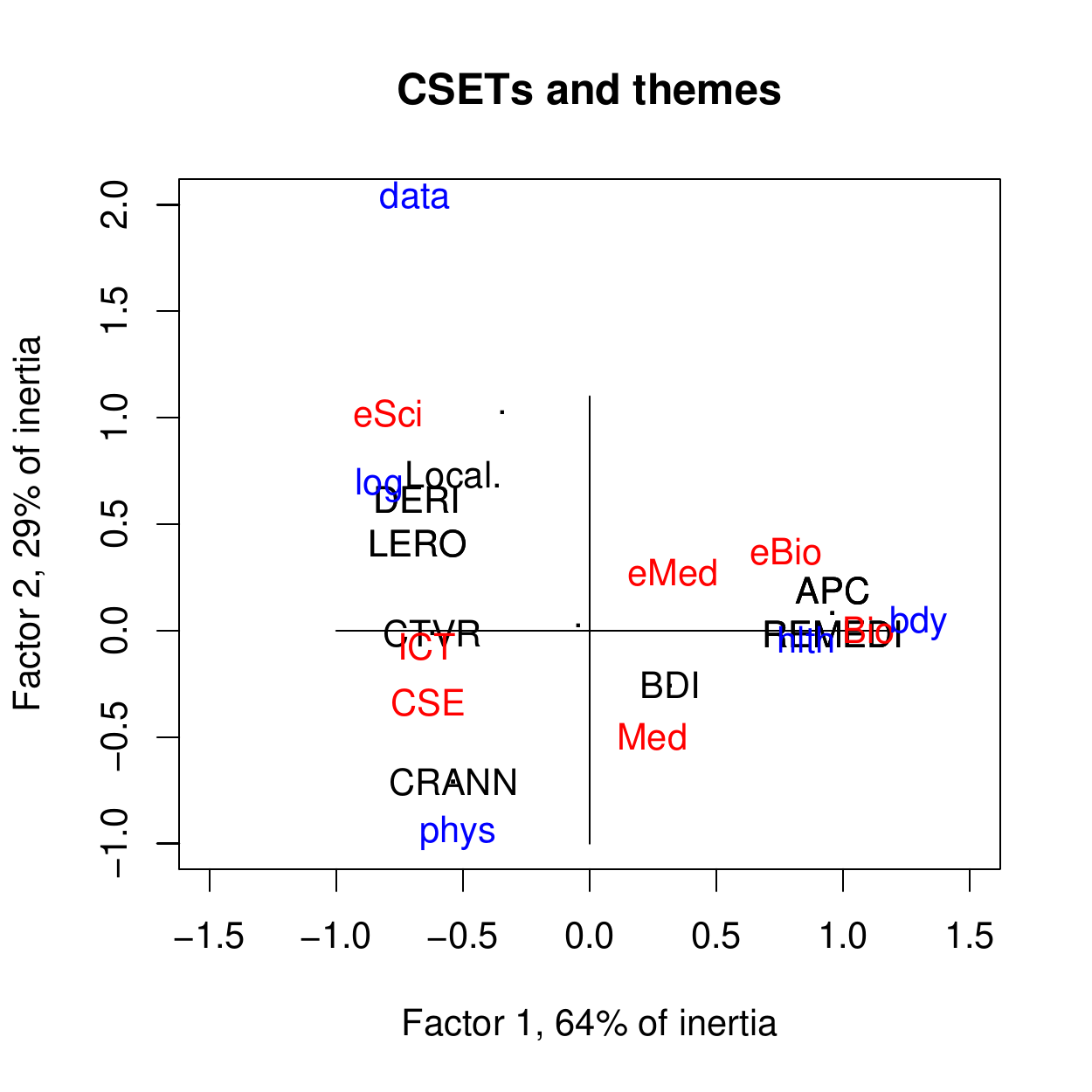}
\end{center}
\caption{As Figure \ref{fig9} but with sub-themes projected into the
display.  Note that, through use of supplementary elements, the axes
and scales are identical to those on Figure \ref{fig9}. 
Axes and scales are just displayed differently in this figure so 
that sub-themes appear in our field of view.}
\label{fig10}
\end{figure}

In Figure \ref{fig9} eight CSETs and major themes are shown.
Factor 1 counterposes computer engineering (left) to biosciences (right). 
Factor 2 counterposes software on the positive end
to hardware on the negative end.  This 2-dimensional map encapsulates 
64\% (for factor 1) + 29\% (for factor 2) = 93\% of all information, i.e.\
inertia, in the dual clouds of points.  
CSETs are positioned relative to the thematic areas used.  In Figure 
\ref{fig10}, sub-themes are additionally projected into the display.  
This is done by taking the sub-themes as 
{\em supplementary elements} following the analysis as such.
From Figure \ref{fig10} we might wish to label additionally factor 2 
as a polarity of data and physics, associated with the extremes of
software and hardware.  

\subsection{Change Over Time}
\label{sect32}

We take another funding programme, the Research Frontiers Programme,
to show how changes over time can be mapped.  

This programme follows an annual call, and includes 
all fields of science, mathematics and engineering.
There are approximately 750 submissions annually.  
There was a 24\% success rate (168 awards) in 2007, and 
19\% (143 awards) in 2008.
The average award was \EUR 155k in 2007, and \EUR 161k in 2008.
An award runs for three years of funding, and this is moving to four 
years in 2009 to accommodate a 4-year PhD duration.
We will look at the Computer Science panel results only, over 2005, 
2006, 2007 and 2008.

Grants awarded in these years, respectively, were: 14, 11, 15, 17.
The breakdown by institutes concerned was: 
UCD -- 13; TCD -- 10; DCU -- 14; UCC -- 6; UL -- 3; 
DIT -- 3; NUIM -- 3; WIT -- 1.  These institutes are as follows: 
UCD, University College Dublin; DCU, Dublin City University; 
UCC, University College Cork; UL, University of Limerick; NUIM, 
National University of Ireland, Maynooth; DIT, Dublin Institute of 
Technology; and WIT, Waterford Institute of Technology.  

One theme was used to characterize each proposal
from among the following: bioinformatics, 
imaging/video, software, networks, data processing \& information 
retrieval, speech \& language processing, virtual spaces, 
language \& text, information security, and e-learning.
Again this categorization of computer science can be contrasted with 
one derived for articles in recent years in the Computer Journal
(Murtagh, 2008).

\begin{figure}
\begin{center}
\includegraphics[width=8cm]{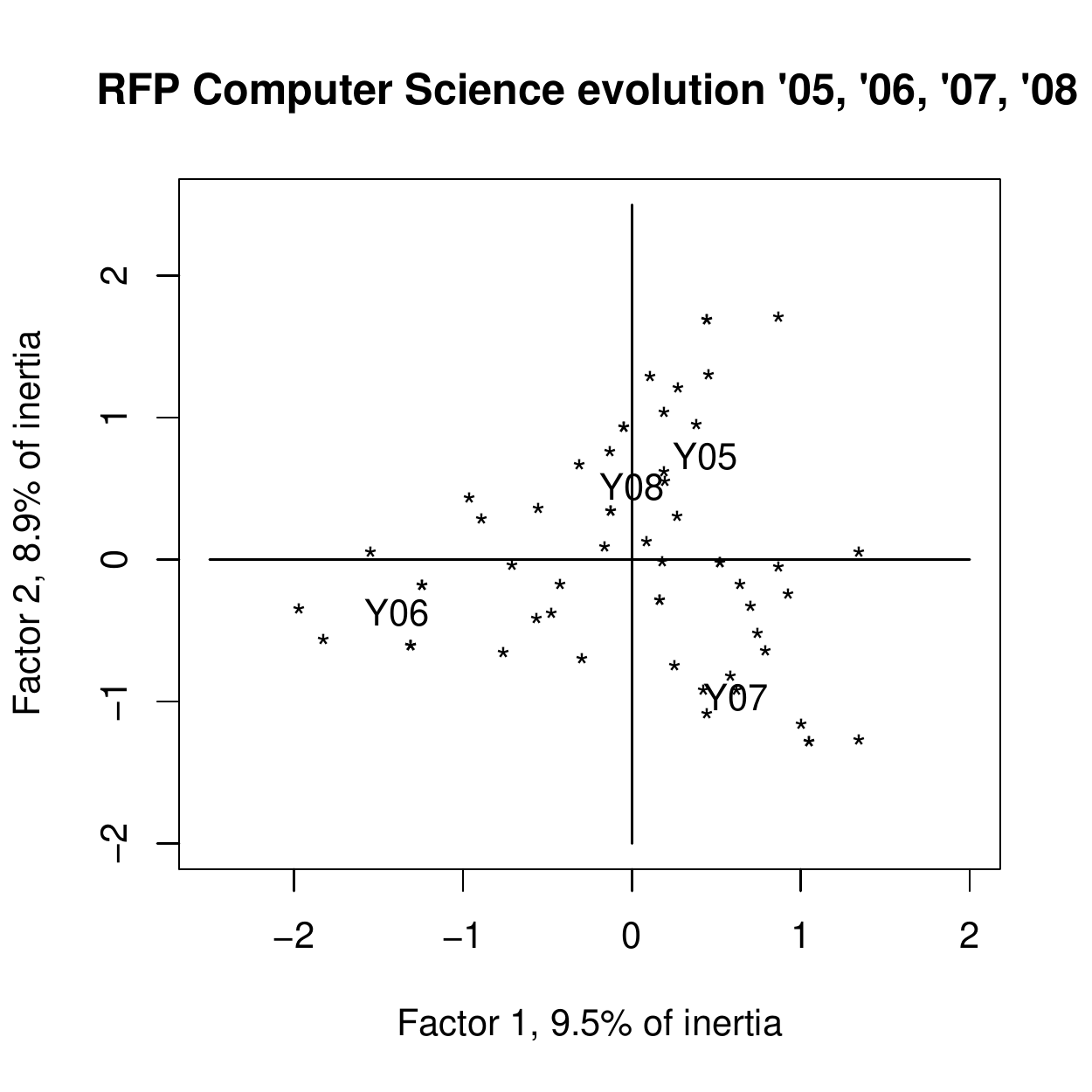}
\end{center}
\caption{Research Frontiers Programme over four years. 
Successful proposals are shown as asterisks.  The years are located 
as the average of successful projects.}
\label{fig14}
\end{figure}

\begin{figure}
\begin{center}
\includegraphics[width=8cm]{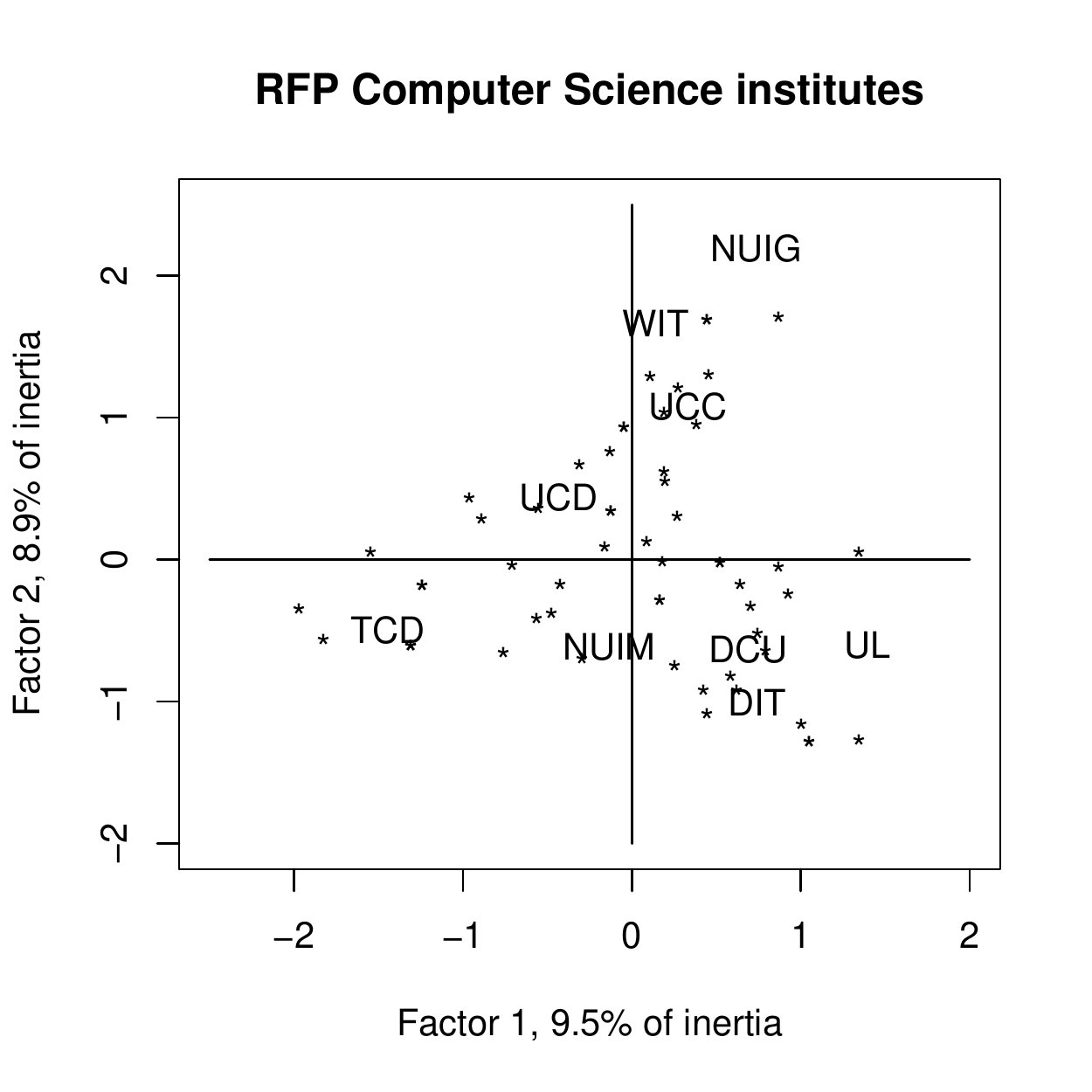}
\end{center}
\caption{As Figure \ref{fig14}, displaying host institutes of the awardees.}
\label{fig15}
\end{figure}

\begin{figure}
\begin{center}
\includegraphics[width=8cm]{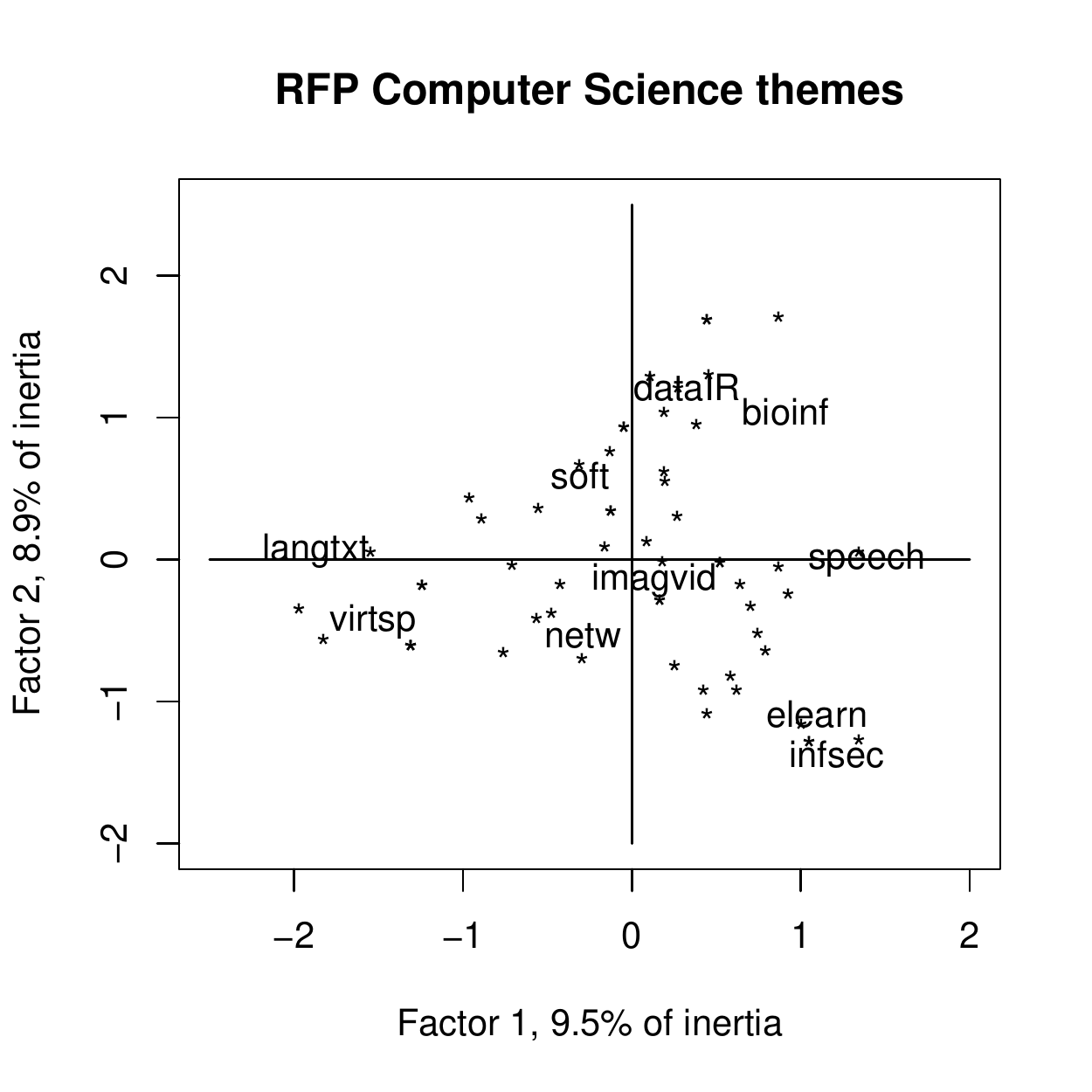}
\end{center}
\caption{As Figures \ref{fig14} and \ref{fig15}, displaying themes.}
\label{fig16}
\end{figure}

Figures \ref{fig14}, \ref{fig15} and \ref{fig16} show different facets of the 
Computer Science outcomes.  By keeping the displays separate, we focus on 
one aspect at a time.  All displays however are based on the same list of 
themes, and so allow mutual comparisons.  Note that the principal 
plane shown accounts for just 9.5\% + 8.9\% of the inertia.  Although 
accounting for 18.4\% of the inertia, this plane, comprising factors, or 
principal axes, 1 and 2, accounts for the highest amount of inertia
(among all possible planar projections).   
Ten themes were used, and what the 18.4\% information content 
tells us is that there is importance attached to most if not all of the ten.

\subsection{Conclusion on the Policy Case Studies}

The aims and objectives in our use of the Correspondence Analysis and 
clustering platform is to drive strategy and its implementation in 
policy.  

What we are targeting is to study highly multivariate, evolving data flows. 
This is in terms of the semantics of the data -- principally, complex webs 
of interrelationships and evolution of relationships over time.
This is the {\em narrative of process} that lies behind raw statistics and 
funding decisions.

We have been concerned especially with {\em information focusing} in 
section \ref{sect31}, and this over time in section \ref{sect32}.

\section{Domain Taxonomy and Researcher's Rank for Data Analysis}
\label{part2}

Here we turn to a domain taxonomy, that is the Computing Classification System 
maintained and updated by the Association of Computing Machinery (ACM-CCS); the latest 
release, of 2012, is publicly available at ACM (2012). 
Parts of ACM-CCS 2012 related to the loosely defined subject of ``data analysis'' 
including ``Machine learning'' and ``Data mining'', up to a rather coarse granularity, 
are presented in Table \ref{t1}. 
\begin{table}[H]
\caption{ACM CCS 2012 high rank items covering data analysis, machine learning and data mining } 
\centering
\begin{tabular}{l|l}
\hline
\multicolumn{1}{c|}{Subject index} & \multicolumn{1}{l}{Subject name} \\
\hline
1.&\quad Theory of computation \\
1.1.&\quad \quad Theory and algorithms for application domains \\
2.&\quad Mathematics of computing \\
2.1.&\quad \quad Probability and statistics \\
3.&\quad Information systems \\
3.1.&\quad \quad Data management systems \\
3.2.&\quad \quad Information systems applications \\
3.3.&\quad \quad World Wide Web \\
3.4.&\quad \quad Information retrieval \\
4.&\quad Human-centered computing \\
4.1.&\quad \quad Visualization \\
5.&\quad Computing methodologies \\
5.1.&\quad \quad Artificial intelligence \\
5.2.&\quad \quad Machine learning \\
\hline
\end{tabular}
\label{t1}
\end{table}

It should be noted that a taxonomy is a hierarchical structure for shaping knowledge. 
The hierarchy involves just one relation ``A is part of B'' so that it leaves aside 
many other aspects of knowledge including, for example, the 
differences between theoretical interrelations, computational issues and application 
matters of the same set of concepts. These, however, may sneak in, even if 
unintentionally, in practice. 
For example, topics representing ``Cluster analysis'' occur in the 
following six branches within the ACM-CCS taxonomy: (i) Theory and algorithms for 
application domains, (ii) Probability and statistics, (iii) Machine learning, (iv) Design 
and analysis of algorithms, (v) Information systems applications, (vi) Information 
retrieval. Among them, (i) and (ii) refer to theoretical work, (iv) to algorithms, (v) 
and (vi) to applications. Item (iii), Machine learning, probably embraces all of them. 

Unlike in biology, the taxonomies of specific research domains cannot be 
specified exactly because of the changing structure of the domain and, therefore, are 
subject to much change.  For example, if one compares  the current ACM Computing 
Classification System 2012 (ACM, 2012) with its previous version, the ACM 
Classification of Computing Subjects 1998 which is available at the same site, one 
cannot help but notice great differences in both the list of sub-domains and the 
structure of their mutual arrangement. 

We consider the set of branches in Table \ref{t1} as a taxonomy of its own, referred 
to below as the Data Analysis Taxonomy (DAT). An extended version of the taxonomy, 
along with three to four more layers of higher granularity, presented in Mirkin and Orlov
(2015, pp.\ 241-249), will be used throughout for illustration of our approach.   

Out of various uses of a domain taxonomy, we pick up here its use for determining a 
scientist rank according to the rank of that node in the taxonomy which has been 
created or significantly transformed because of the results by the scientist 
(Mirkin, 2013).
  
The concept of taxonomic rank is not uncommon in the sciences. It is quite popular, for 
example, in biology: ``A Taxonomic Rank is the level that an 
organism is placed within the hierarchical level arrangement of life forms'' (see 
http://carm.org/dictionary-taxonomic-rank). As mentioned in Mirkin and Orlov (2015), {\it Eucaryota} 
is a domain (rank 1) containing {\it Animals} kingdom (rank 2). The latter contains 
{\it Cordata} phylum (rank 3) which contains {\it Mammals} class (rank 4) which 
contains {\it Primates} order (rank 5) which contains {\it Hominidae} family 
(rank 6) which contains {\it Homo} genus (rank 7) which contains, finally, 
{\it Homo sapiens} species (rank 8). Similarly, the rank of the scientist who created 
the ``World wide web'' (Berners-Lee, 2010), (the item 3.3 in Table \ref{t1}) at layer 2 of the 
DAT taxonomy, is 2; and the rank of the scientist who developed a sound theory for 
``Boosting'' (Schapire, 1990), (the item 1.1.1.5 in DAT (Mirkin and Orlov, 2015)), is 4, whereas the 
rank of the scientists who proposed a sound approach to ``Topic modeling'' (Blei et al., 2003)
(the item 5.2.1.2.4 in DAT (Mirkin and Orlov, 2015)) is 5.
This specification of taxonomic rank, TR, is associated with qualitative innovation, 
whereas
the dominant current approach is to only reward or take account of low rank, and 
particular, topic items.   

Using taxonomic ranks (TRs) based on domain taxonomies for 
evaluating the quality of research differs from the other methods currently 
available, through the following features:
\begin{itemize}
\item The TR method directly measures the quality of results themselves rather than any 
related feature such as popularity;
\item The TR evaluation is well subject-focused; a scientist with good results in 
optimization may get rather modest evaluation in data analysis because a taxonomy 
for data analysis would not include high-level nodes on optimization;
\item The TR rank can get reversed if the taxonomy is modified so that the rank-giving 
taxon gets a less favourable location in the hierarchical tree;
\item The granularity of evaluation can be changed by increasing the granularity of the 
underlying taxonomy;
\item The TR evaluations in different domains can be made comparable by using taxonomies 
of the same depth; 
\item The maintenance of a domain taxonomy can be effectively organized by a research 
community as a special activity subject to regular checking and scrutinising;
\item Assigning the TR to a scientist or their result(s) is derived from mapping them 
to a sub-domain that has been significantly affected by them, and this is not a simple 
issue. The persons who do the mapping must be impartial and have deep knowledge of the 
domain and the results. 
\end{itemize}
 
The last two items in the list above refer to the core of the proposal in this paper. 
They can be considered  a clarification of the main claim  over evaluation of the 
research impact made by the scientists: qualitative considerations should prevail 
over metrics (Dora, 2013; Hicks et al., 2015; Metric Tide, 2016). 
Here the wide meaning of ``qualitative'' is 
reduced to two points: (a) developing and maintaining of a taxonomy, and (b) mapping 
results to the taxonomy. Both taxonomy developing any mapping decisions involve 
explicitly stated judgements which can be discussed openly and corrected if needed.  
This differs greatly from the currently employed procedures of peer-reviewing which 
can be highly subjective and dependent on various external considerations 
(Eisen et al., 2013; Engels et al., 2013; Van Raan, 2006). 
The activity of developing and maintaining taxonomies can be 
left to the governmental agencies and funding bodies, or to scholarly academies, 
or to discipline and sub-discipline expert organisational bodies, whereas the mapping 
activity should be left, in a transparent way, to scientific discussions involving all 
relevant individuals. 
Of course, there is potential for further developments of the formats: taxonomies 
could be extended to include various aspects characterizing research developments, 
and mapping can be softened up to include spontaneous and uncertain judgements.  

\section{A Prototype of Empirical Testing} 

We focus on the field of Computer Science related to data analysis, machine learning, 
cluster analysis and data mining along with its taxonomy derived from the ACM Computing 
Classification System 2012 (ACM, 2012), as explained above. We pick up a sample of 30 
leading scientists in the field (about half from the USA, and other, mostly European, 
countries are represented by 2--3 representatives), such that the information of their 
research results is publicly available. Although we tried to predict the leaders, their 
Google-based citation indexes are highly different, from a few thousand to a hundred 
thousand. We picked up 4--6 most important papers by each of the sampled scientists 
and manually mapped each of the papers to taxons significantly affected by that. Since 
some of the relevant subjects, such as ``Consensus clustering'' and ``Anomaly detection'', 
have not been presented in the ACM-CCS, we added them to DAT (Data Analysis Taxonomy) as 
leaves, implying that a previous terminal node becomes a non-terminal node. 
The results of the mapping  are presented in Table \ref{t3}. The table also presents 
the derived taxonomic ranks and the same ranks, 0--100 normalized.
To derive the taxonomic rank of a scientist, we first take the minimum of their ranks 
as the base rank. Then we subtract from it as many one tenths as there are subdomains 
of that rank in their list and as many one hundredths as there are subdomains of greater 
ranks in the list. For example, the list of S23 comprises ranks 4, 5, 4 leading to 4 as 
the base rank. Subtraction of two tenths and one hundredth from 4 gives the derived 
rank 3.79.
 The normalization is such that the minimum rank, 3.50, gets a 100 mark, and the maximum 
rank, 4.89, gets a 0. The last column, the stratum, is assigned according to the 
distance of the mark to either 70 or 30 or 0.

\setlength{\tabcolsep}{6pt}

\begin{table}[H] \small
\vspace{-3pt}
\caption{Mapping main research results to the taxonomy; layers of the nodes affected; 
Tr -- taxonomic ranks derived from them; Trn -- taxonomic ranks normalized to the range 
0 to 100; and three strata obtained by k-means partitioning of the ranks. } 
\centering
\begin{tabular}{c|c|c|*{2}{r|}c}  \hline 
Scientist & Mapping to taxonomy & Layers & Tr & Trn & Stratum \\ \hline \hline
S1 &  4.1.2.7, 5.2.1.2.7, 5.2.3.7.7   & 4,5,5 & 3.88 & 73 & 1 \\ \hline
S2 &  2.1.1.2, 2.1.1.2,  5.2.2.7, 5.2.3.5,  5.2.3.5 & 4,4,4,4,4 & 3.50 & 100 & 1 \\ \hline
S3 &   3.2.1.4.2, 5.2.1.2.3, 5.2.1.2.7, 5.2.3.5.4 ,& & & & \\
&  5.2.3.7.6 & 5,5,5,5,5 & 4.50 & 29 & 2 \\ \hline
S4 &  1.1.1.4.3, 3.4.4.5, 5.2.1.1.1,5.2.1.2.7, & &  &  & \\
  &  5.2.3.2.1,5.2.3.7.8  & 5,4,5,5,5,5 & 3.90 & 71 & 1 \\ \hline
S5 &  3.2.1.4.4, 3.2.1.4.4, 3.2.1.4.5, 3.2.1.4.6, 3.2.1.11.1 & 5,5,5,5,5 & 4.50 & 29 & 2 \\ \hline
S6 &  3.1.1.5.2, 3.1.2.1.1, 3.1.2.1.1 , & & & &\\ 
  &  3.2.1.6., 3.2.1.7& 5,5,5,4,4 & 3.77 & 81 & 1 \\ \hline
S7 &  5.2.3.5.6,  5.2.3.5.7 & 5,5 & 4.80 & 7 & 3 \\ \hline
S8 &  3.2.1.3.1, 3.2.1.4.1, 5.2.3.3.1, 5.1.3.2.1, 5.1.3.2.4 & 5,5,5,5,5 & 4.50 & 29 & 2 \\ \hline
S9 &  5.2.1.2.3, 5.2.3.3.2, 5.2.3.5.1, 5.2.3.5.4, & & & & \\ 
  & 5.2.3.6.2 & 5,5,5,5,5 & 4.50 & 29 & 2 \\ \hline
S10 & 5.2.3.3.2,  5.2.3.13.1  & 5,5 & 4.80 & 7 & 3 \\ \hline
S11 & 3.2.1.2, 3.2.1.2.1,3.2.1.3.3,3.2.1.4.1, &  &  &  & \\
  & 3.2.1.7.2 & 4,5,5,5,5 & 3.86 & 74 & 1 \\ \hline
S12 & 3.2.1.9.1.1,3.2.1.10,3.2.1.11.2,5.1.1.7.1, &  &  &  &  \\
  &  5.2.3.1.3,5.2.3.4.1  & 6,4,5,5,5,5 & 3.86 & 74 & 1 \\ \hline
S13 & 1.1.1.3, 5.2.1.2.1,5.2.1.2.1,5.2.2.7.1, &  &  &  &\\
  &  5.2.3.7.1 & 4,5,5,5,5 & 3.86 & 74 & 1 \\ \hline
S14 &  3.2.1.3.1 & 5 & 4.90 & 0 & 3 \\ \hline
S15 &  5.2.4.3.1 & 5 & 4.90 & 0 & 3 \\ \hline
S16 &  5.2.4.2.3 & 5 & 4.90 & 0 & 3 \\ \hline
S17 &  2.1.3.7.1, 5.2.4.3.1, 5.2.3.7.5., 5.2.1.2.4, &  &  &  &\\ 
  &  5.2.3.2.4, 5.2.3.7.3.2, 5.2.3.5.4., 5.2.4.3.1 & 5,5,5,5,6,5,5 & 4.39 & 36 & 2 \\ \hline
S18 &  3.2.1.9.1,3.2.1.9.2,5.2.3.3.3.1 & 5,5,6 & 4.79 & 8 & 3 \\ \hline
S19 &  3.2.1.7.5,3.2.1.9.3,5.2.3.2.1.1,5.2.4.5.1 & 5,5,6,5 & 4.69 & 15 & 3 \\ \hline
S20 &  3.2.1.4.3,5.2.3.7.7,5.2.3.7.8.1 & 5,5,6 & 4.79 & 8 & 3 \\ \hline
S21 &  1.1.1.6,2.1.1.2,2.1.1.8.3,3.2.1.6, &  &  &  & \\ 
  &  3.4.1.6,5.1.2.4,5.2.1.1.3& 4,4,5,4,4,4,5 & 3.57 & 95 & 1 \\ \hline
S22 &  3.2.1.2.2,5.2.1.2.7.1,5.2.3.1.2,5.2.3.6.2.1& 5,6,5,6 & 4.78 & 9 & 3 \\ \hline
S23 &  3.2.1.3,3.2.1.3.1,3.4.4.1 & 4,5,4 & 3.79 & 79 & 1 \\ \hline
S24 &  2.1.5.3.1 & 5 & 4.90 & 0 & 3 \\ \hline
S25 &  5.2.3.3.3.2, 5.2.3.8.1 & 6,5 & 4.89 & 1 & 3 \\ \hline
S26 &  3.2.1.11.1,3.2.1.11.1,3.3.1.6,5.2.2.7, &  &  &  &  \\
  & 5.2.3.5.6& 5,5,4,4,5 & 3.77 & 81 & 1 \\ \hline
S27 &  3.2.1.3.2,3.2.1.4.1,5.2.1.2.1,5.2.3.1.1& 5,5,5,5 & 4.60 & 21 & 2 \\ \hline
S28 &  3.2.1.8 & 4 & 3.90 & 71 & 1 \\ \hline
S29 &  5.2.3.3.2.1,5.2.3.3.3.3,5.2.3.3.4& 6,6,5 & 4.88 & 1 & 3 \\ \hline
S30 &  5.1.3.2.1.1,5.2.1.2.7.2,5.2.3.3.5& 6,6,5 & 4.88 & 1 & 3 \\ \hline
\end{tabular}
\label{t3}
\vspace{-2pt}
\end{table}

\vspace{-5pt}
\section{Comparing Taxonomic Rank with Citation and Merit}

We compared our taxonomic ranks with more conventional criteria: (a) Citation and (b)  
Merit. The Citation criterion was derived from Google-based indexes of the total number 
of citations, the number of works receiving 10 or more citations, and Hirsch index h, 
the number of papers receiving h citations or more. The merit criterion was computed 
from data on the following three indices: the number of successful PhDs (co)-supervised, 
the number of conferences co-organized, and the number of journals for which the
researcher-scientist is a member of the Editorial Board.

To aggregate the indexes into a convex combination, that is, a weighted sum, 
automatically, 
a principled approach which works for correlated or inconsistent criteria has been 
developed. According to this approach, given the number of strata (in our case 3), the 
aggregate criterion is to be found so that its direction in the index space is such that 
all the observations  are projected into compact well-separated clusters along this 
direction (Mirkin and Orlov, 2013, 2015).

To be more specific, consider a data matrix scientist-to-criteria $\mathbf{X}=(x_{ij})$ where 
$i=1,..., N$ are indices of scientists, $j=1,...,M$ are indices of $M$ criteria, and 
$x_{ij}$  is the score of $j$–th criterion for the $i$–th scientist. Let us consider a 
weight vector $\mathbf{w}=(w_1, w_2,..., w_M)$ such that $w_j\ge 0$ for every $j$ and $\sum_j 
w_j=1$, for the set of criteria. 
Then  the combined criterion is $\mathbf{f}=\sum_{j=1}^M w_jx_j$ where $x_j$ is $j$th column 
of matrix $\mathbf{X}$. The problem is to find $K$ disjoint subsets $S=\{S_1,.. S_k, ..., S_K\}, 
k=1,...,K$ of the set of
 indices $i$, referred to as strata, according to values of the combined criterion $\mathbf{f}$. 
Each stratum $k$ is characterized by a value of the combined criterion $c_k$,  the 
stratum  centre. Geometrically, strata are formed by layers between parallel planes 
in the space of criteria. At any stratum  $S_k$, we want the value of the combined 
criterion 
$f_i=\sum_{j=1}^M w_j x_{ij}$ at any $i\in S_k$ to approximate the stratum centre 
$c_k$. In other words, 
 in  the equations $x_{i1}w_{1}+x_{i2}w_{2}+...+x_{iM}w_{M}=c_{k} + e_{i}$,  $e_{i}$ 
are errors to be minimized over vector $\mathbf{w}$. A least-squares formulation of the linear 
stratification (LS) problem: find a vector  $\mathbf{w}$, a set of centres $\{c\}$ and a 
partition $S$ to solve the problem in (\ref{eq1}), as follows.

\begin{equation}
\begin{aligned}
& \underset{w,c,S} {\text{min}}
& &\sum\limits_{k=1}^{K} \sum\limits_{i \in S_k} (\sum\limits_{j=1}^{M} x_{ij}w_j-c_k)^2  \\
& \text{such that} & &  \sum\limits_{j=1}^{M} w_j = 1 \\
& & & w_j\ge 0, j \in {1...M}. \label{eq1}
\end{aligned}
\end{equation}

This problem can be tackled using the alternating minimzation approach, conventional 
in cluster analysis.  For any given weight vector $\mathbf{w}$, the criterion in (\ref{eq1}) is 
just the conventional square-error clustering criterion of the $K$-means clustering 
algorithm over a single feature, the combined criterion $\mathbf{f}=\sum_{j=1}^M w_jx_j$.  
Finding an appropriate $\mathbf{w}$ at a given stratification $S$ can be reached by using 
standard quadratic optimization software. 

\begin{figure}  
\includegraphics[scale=0.4]{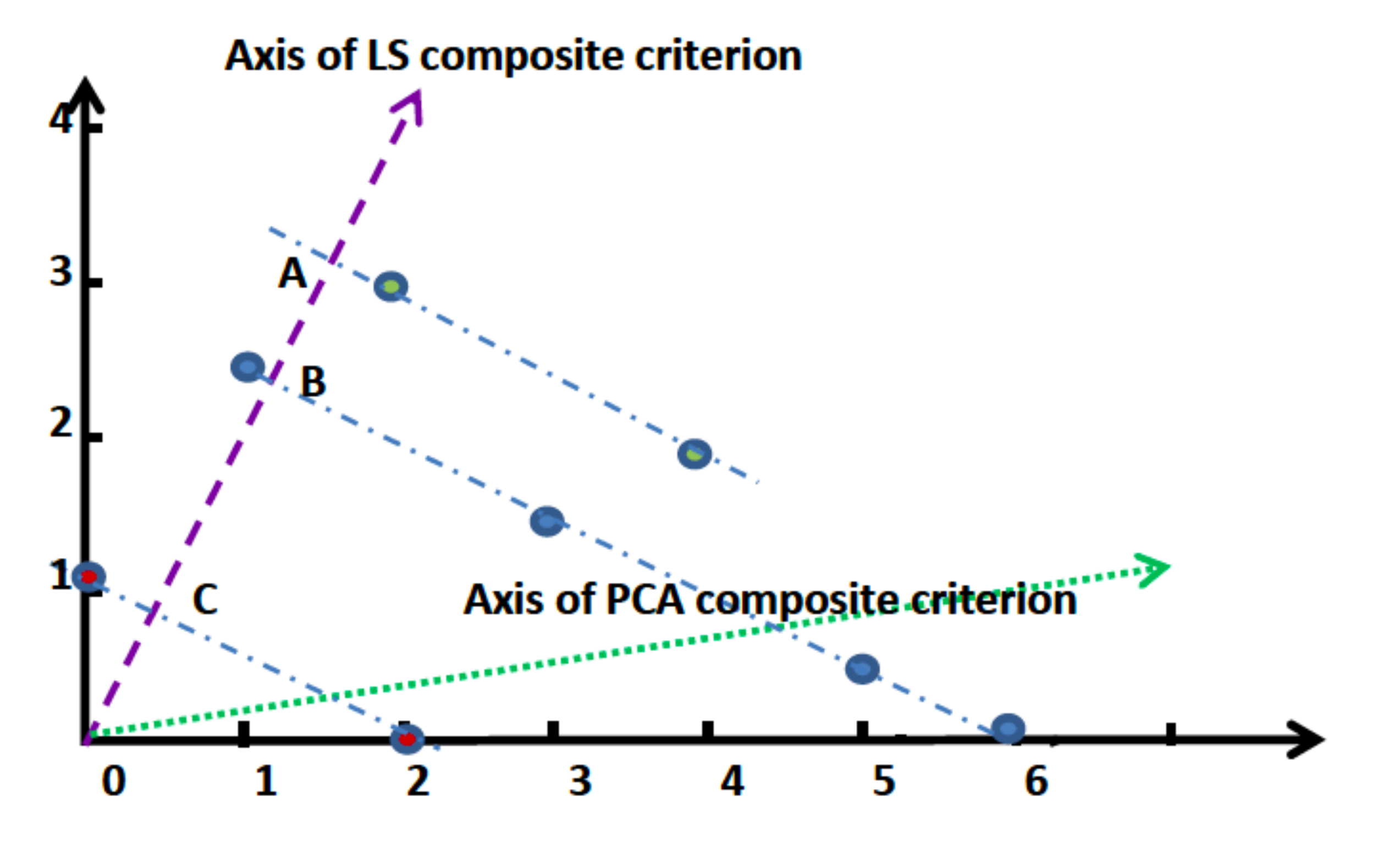}
\caption{Eight scientists on the plane of criteria $x$ and $y$. The LS and PCA 
combined criteria are represented with broken lines.}
\label{fm}
\end{figure}

To illustrate the approach as it is and, also, its difference from the widely used Principal 
Component Analysis (PCA) approach to linearly combining criteria, let us consider the 
following example.
In Table \ref{tm1}, scores of two criteria over 8 scientists are presented.

\begin{table} 
\caption{Scores of two criteria, $x$ and $y$, over 8 scientists 
 labelled, for convenience, by using an uppercase notation of the 
corresponding strata (see Figure \ref{fm}).} 
\centering
\begin{tabular}{l|c|c}
\hline
Label & Criterion $x$ & Criterion $y$\\
\hline
C1&    2 &    0\\
C2 &   0 &   1\\
B1 &   6 &   0\\
B2 &  5  &  0.5\\
B3&    3  &  1.5\\
B4&    1  &  2.5\\
A1&    4   & 2\\
A2&    2  &   3\\
\hline
\end{tabular}
\label{tm1}
\end{table}

Although usually criteria values are normalized into a 0--100\% scale, we do not do 
that here to keep things simple.
It appears, the data ideally, with zero error, fall into three strata, $K=3$, as shown in
Figure \ref{fm}, according to combined criterion $\mathbf{f}=\frac{1}{3}x+\frac{2}{3}y$. In 
contrast, the PCA based 
linear combination, $\mathbf{z}=  0.7712 \mathbf{x} + 0.2288 \mathbf{y}$, admits a residual of 13.4\% of the total 
data scatter, and leads to a somewhat
different ordering, at which two top stratum scientists get lesser aggregate scores 
than two scientists of the B stratum. 

For convenience, the combined criteria scores are presented in Table \ref{tm2}.
\begin{table}[!]
\caption{Scores of two combined criteria, the LS based and PCA based.}
\label{tm2}
\centering
\begin{tabular}{l|c|c}
\hline
Label& LS & PCA\\
\hline
C1&  0.67 &  1.54\\
C2 &   0.67 &   0.23\\
B1 &   2.00 &   4.63\\
B2 &  2.00 &  3.97\\
B3&   2.00  &  2.66\\
B4&   2.00 &  1.34\\
A1&    2.67   & 3.54\\
A2&    2.67 &   2.23\\
\hline
\end{tabular}
\end{table}

In the thus aggregated Citation criterion, the Hirsch index received a zero coefficient, 
while the other two were one half each.  The zeroing of the Hirsch index weight is in 
line with the overwhelming critiques this index has been exposed to in recent times,
(Albert, 2013; Osterloh and Frey, 2014; Dora, 2013; Van Raan, 2006).
A similarly aggregated Merit criterion  is formed with weights
  0.22 for the number of PhD students, 0.10 for the number of conferences, and 0.69 for
the number of journals, which is consistent with the prevailing practice of maintaining 
a heavy and just submission reviewing process in leading journals. 

To compare these scales, let us compute Pearson correlation coefficients between them, 
see Table \ref{t08}.

\begin{table}[H]
\caption{Pairwise correlations between criteria (only the part above the diagonal is 
shown).} 
\centering
\begin{tabular}{lcc}
\hline
 Criterion  &    Citation & Merit \\
\hline
TR             &  -0.12 & -0.04   \\
Citation       &        & 0.31\\
\hline
\end{tabular}
\label{t08}
\end{table}

As expected, the Citation and Merit criteria do not correlate with the Taxonomic rank of 
the scientists. On the other hand, the traditional  Citation and Merit criteria are 
somewhat positively correlated, probably because they both relate to the popularity of 
a scientist.

\section{Conclusions}
Assessments can be carried out at different levels, a region, an organization, a team or an individual researcher;
within a domain or inter domains. 
What we can metaphorically express as wider horizons, are brought to our 
attention, through analysis of quality.  Among the recommendations arising from
this work, on the regional level, there are three on the particular subjects of our concern:
 
\begin{itemize}
\item Set out a more structured and strategic process for proposing projects. 

\item Conduct a systematic analysis of the existing infrastructure.

\item Take a more systematic approach to evaluating the impact of operational 
projects.
\end{itemize}

With these recommendations, we are emphasizing the importance of these underpinning
themes.  These themes, and their underpinnings, should be pursued assertively 
for journals and other scholarly 
publishing, and  also for research funding programmes.  

We both observe and demonstrate that evaluation of research, especially at the level of teams or individuals  can be organized by,
firstly, developing 
and maintaining a taxonomy of the relevant subdomains and, secondly, a system for mapping research 
results to those 
subdomains that have been created or significantly transformed because of these 
research results. This 
would bring a well-defined meaning to the widely-held opinion that 
research impact should be evaluated, first of all, based on qualitative considerations. 
Further steps can be, and should be, undertaken
in the directions of developing and maintaining a system for assessment of the 
quality of research across all areas of knowledge. 
Of course, developing and/or incorporating systems for other elements of research 
impact, viz., knowledge transfer, industrial applications, social interactions,  etc., 
are to be taken into account also. 
In comprehensively covering quality and quantitative research outcomes, 
there can be distinguished at least five aspects of an individual researcher's research impact:
\begin{itemize}
\item Research and presentation of results (number, quality)
\item Research functioning (journal/volume editing, running research meetings, reviewing)
\item Teaching (knowledge transfer, knowledge discovery)
\item Technology innovations (programs, patents, consulting)
\item  Societal interactions (popularization, getting feedback)
\end{itemize}

Many, if not all, of the items in this list can be maintained by developing and using corresponding taxonomies.
The development of a system of taxonomies for the health system in the USA, IHTSDO SNOMED CT (SNOMED CT, 2016),
extended now to many other countries, and languages, should be considered an instructive example of such a major undertaking.

This suggests directions for future work. Among them are the following.

In methods: (i) Enhancing the concept of taxonomy by including theoretical, computational, and
industrial facets, as well as dynamic aspects to it; (ii) Developing methods for 
relating paper's texts, viz.\ content, and taxonomies; (iii) Developing methods for 
taxonomy building using such research paper texts, i.e.\ content; (iv) Developing 
methods for mapping research results to taxonomy units affected by them; (v) 
 Using our prototyping here, developing comprehensive methods for ranking the impact 
of results to include expert-driven components; (vi) Also based on our prototyping here, developing accessible and 
widely used methods for aggregate rankings.

In substance: (i) Developing and maintaining a permanent system for assessment of 
the scope and quality 
of research at different levels; (ii) Developing a system of domains in research subjects and their taxonomies;
(iii) Cataloguing researchers, research and funding bodies, and research results; (iv) Creating a platform and 
forums for discussing taxonomies, results and assessments.

A spin-off of our very major motivation for qualitative analytics is to propose using a full 
potential of the research efforts on a regional level. 
In our journal editorial roles, we realise very well that sometimes quite predictable
rejection of article submissions can raise such 
questions as the following: is there no qualitative interest at all in such work?
How can, or how should, improvement be recommended?  At least as important, and 
far more so in terms of wasteful energy and effort, is the qualitative analysis 
of rejected research funding proposals.  (As is well known, a relatively small proportion of the research projects
gets a ``go ahead'' nod. For example,  The European Horizon 2020 
FET-Open, Future Emerging Technologies, September 2015 proposal submission 
 resulted in less than a 2\% success rate (FET, 2016): 13 successful research 
proposals out of 822 proposal submissions.)  Given the workload at issue, 
on various levels and from various vantage points, there is potential for 
data mining and knowledge discovery in the vast numbers of rejected research funding
proposals.  Ultimately, and given the workload 
undertaken, it is both potentially of benefit, and justified, to carry out such 
analytics.

\section*{References}

\smallskip

\noindent
ABRAMO, G., CICERO, T., ANGELO, C.A. (2013).
``National peer-review research assessment exercises for the hard sciences can be a 
complete waste of money: the Italian case'',  {\em Scientometrics}, 95(1), 311--324.

\smallskip

\noindent
ACM (2012). 
The 2012 ACM Computing Classification System, http://www.acm.org/about/class/2012
(Viewed 2017-02-05).

\smallskip

\noindent 
ALBERT, B. (2013). ``Impact factor distortions'', {\em Science}, 340, no. 6134, 787.

\smallskip

\noindent
ARAGN\'ON, A.M. (2013).  ``A measure for the impact of research'',
{\em Scientific Reports} 3, Article number: 1649.

\smallskip

\noindent
BERNERS-LEE, T. (2010). ``Long live the Web'', {\em Scientific American}. 303 (6). 80--85.

\smallskip

\noindent
BLEI, D.M., NG, A.Y., JORDAN, M.I., LAFFERTY, J. (2003). ``Latent Dirichlet allocation'', 
{\em Journal of Machine Learning Research}. 3: 993--1022.

\smallskip

\noindent
CANAVAN, J., GILLEN, A., SHAW, A. (2009).  ``Measuring research impact: developing 
practical and cost-effective approaches'', 
{\em Evidence and Policy: A Journal of Research, Debate and Practice}, 5.2. 167--177.

\smallskip

\noindent
DORA (2013).  
San Francisco Declaration on Research Assessment (DORA), www.ascb.org/dora
(viewed 2017-02-05).

\smallskip

\noindent
EISEN, J.A., MACCALLUM, C.J., NEYLON, C.  (2013).  
``Expert failure: Re-evaluating research assessment''. 
{\em PLoS Biology}, 11(10): e1001677.

\smallskip

\noindent
ENGELS, T.C., GOOS, P., DEXTERS, N., SPRUYT, E.H. (2013).  
``Group size, h-index, and efficiency in publishing in top journals explain expert panel 
assessments of research group quality and productivity''. 
{\em Research Evaluation}, 22(4), 224--236.

\smallskip

\noindent
FET (2016), 
``FET-Open: 3 new proposals start preparation for Grant Agreements'', 
{\em Future Emerging Technologies Newsletter}, 21 March 2016. \\
http://ec.europa.eu/newsroom/dae/itemdetail.cfm?item\_id=29587\&newsletter=129

\smallskip

\noindent
HICKS, D., WOUTERS, P., WALTMAN, L., DE RIJCKE, S., RAFULS, I. (2015).
``The Leiden Manifesto for research metrics''. {\em Nature}, 520, 429--431.  

\smallskip

\noindent
SNOMED CT (2016). 
IHTSDO, International Health Terminology Standards Development Organisation, 
SNOMED CT, Systematized Nomenclature of Medicine, Clinical Terms.  \\
http://www.ihtsdo.org/snomed-ct
(viewed 2017-02-05).

\smallskip

\noindent
LEE, F.S., PHAM, X., GU, G. (2013). 
``The UK research assessment exercise and the narrowing of UK economics''. 
{\em Cambridge Journal of Economics}, 37(4), 693--717.

\smallskip

\noindent
METRIC TIDE (2016).  
``The Metric Tide: Report of the Independent Review of the Role of Metrics in Research Assessment and Management'', \\
http://www.hefce.ac.uk/pubs/rereports/Year/2015/metrictide/Title,104463,en.html
(viewed 2017-02-05).

\smallskip

\noindent
MIRKIN, B. (2013).  
``On the notion of research impact and its measurement'', {\em Control in 
Large Systems, Special Issue: Scientometry and Experts in Managing Science}. 
44. 292--307, Institute of Control Problems, Moscow (in Russian). 

\smallskip

\noindent
MIRKIN, B., ORLOV, M. (2013). ``Methods for Multicriteria Stratification and  
Experimental Comparison of Them'', 
Preprint WP7/2013/06, Higher School of Economics, Moscow, 31 pp. (in Russian).

\smallskip

\noindent
MIRKIN, B., ORLOV, M. (2015). ``Three aspects of the research impact by a 
scientist: measurement methods and an empirical evaluation'', 
in A. Migdalas, A. Karakitsiou, Eds., 
{\em Optimization, Control, and Applications in the Information Age}, 
Springer Proceedings in Mathematics and Statistics. 130. 233--260.

\smallskip

\noindent
MURTAGH, F. (2008). ``Editorial''. {\em The Computer Journal}, 51(6), 612--614. 

\smallskip

\noindent
MURTAGH, F. (2010).  
``The Correspondence Analysis platform for uncovering deep structure 
in data and information''. {\em The Computer Journal}, 53(3), 304--315.

\smallskip

\noindent
NG, W.L. (2007).  
``A simple classifier for multiple criteria ABC analysis''. 
{\em European Journal of Operational Research}. 177. 344--353.

\smallskip

\noindent
ORLOV, M., MIRKIN, B. (2014). 
``A concept of multicriteria stratification: A definition and solution'', 
{\em Procedia Computer Science}, 31, 273--280.

\smallskip

\noindent
OSTERLOH, M., FREY, B.S. (2014).  ``Ranking games''. {\em Evaluation review}, 
Sage, pp.\ 1--28.

\smallskip

\noindent
RAMANATHAN, R. (2006). 
``Inventory classification with multiple criteria using weighted linear optimization'', 
{\em Computers and Operations Research}. 33.  695--700.

\smallskip

\noindent
SCHAPIRE, R.E. (1990).  ``The strength of weak learnability''. 
{\em Machine Learning}. 5(2), 197--227.

\smallskip

\noindent
SIDIROPOULOS, A., KATSAROS, D., MANOLOPOULOS, Y. (2014). 
``Identification of influential scientists vs.\ mass producers by the perfectionism 
index''.  Preprint, ArXiv:1409.6099v1, 27 pp.  

\smallskip

\noindent
SUN, Y., HAN, J., ZHAO, P., YIN, Z., CHENG, H., WU, T. (2009).  
``RankClus: integrating clustering with ranking for heterogeneous information 
network analysis''. 
{\em EDBT '09 Proceedings of the 12th International Conference on Extending 
Database Technology: Advances in Database Technology}, ACM, NY, 565--576.

\smallskip

\noindent
THOMSON REUTERS (2016). 
``Thomson Reuters intellectual property and science''. 
(Acquisition of the Thomson Reuters Intellectual Property and Science Business 
by Onex and Baring Asia Completed.
Independent business becomes Clarivate Analytics)
http://ip.thomsonreuters.com
(Viewed 2017-02-05).

\smallskip

\noindent
UNIVERSITY GUIDE (2016). 
``The Complete University League Guide''.    \\
http://www.thecompleteuniversityguide.co.uk/league-tables/methodology.
(Viewed 2017-02-05)

\smallskip

\noindent
VAN RAAN, A.F. (2006).  
``Comparison of the Hirsch-index with standard bibliometric indicators 
and with peer judgment for 147 chemistry research groups''. 
{\em Scientometrics}, 67(3), 491--502.

\end{document}